\documentclass[12pt]{article}
\usepackage{amssymb,amsfonts}
\usepackage{graphics,psboxit,amsmath}
\usepackage{subfigure}
\usepackage{graphicx}
\usepackage{verbatim}
\usepackage{color}
\usepackage{hyperref}
\hypersetup{colorlinks}

\definecolor{darkred}{rgb}{1,0,0}
\definecolor{darkgreen}{rgb}{0,0.5,0}
\definecolor{darkblue}{rgb}{0,0,1}
\definecolor{orange}{rgb}{1,0.5,0}
\definecolor{green}{rgb}{0,1,0}
\definecolor{purple}{rgb}{.5,0,1}

\hypersetup{ colorlinks,
linkcolor=darkred,
filecolor=darkgreen,
urlcolor=darkblue,
citecolor=darkblue,
linktocpage=true }

\definecolor{markcolor}{rgb}{.25,0,1}

\definecolor{markcolor2}{rgb}{1,0,0}

\definecolor{markcolor3}{rgb}{0,1,0}

\def\hybrid{\topmargin -10pt    \oddsidemargin 0.1in 
        \headheight 0pt \headsep 0pt
        \textwidth 16.2cm      
        \textheight 22.2cm       
        \marginparwidth .875in
        \parskip 5pt plus 1pt   \jot = 1.5ex}

\hybrid

\def\baselinestretch{1.2}

\catcode`\@=11

\def\marginnote#1{}
\newcount\hour
\newcount\minute
\newtoks\amorpm
\hour=\time\divide\hour by60 \minute=\time{\multiply\hour by60
\global\advance\minute by-\hour}
\edef\standardtime{{\ifnum\hour<12 \global\amorpm={am}%
        \else\global\amorpm={pm}\advance\hour by-12 \fi
        \ifnum\hour=0 \hour=12 \fi
        \number\hour:\ifnum\minute<10 0\fi\number\minute\the\amorpm}}
\edef\militarytime{\number\hour:\ifnum\minute<10 0\fi\number\minute}

\def\draftlabel#1{{\@bsphack\if@filesw {\let\thepage\relax
   \xdef\@gtempa{\write\@auxout{\string
      \newlabel{#1}{{\@currentlabel}{\thepage}}}}}\@gtempa
   \if@nobreak \ifvmode\nobreak\fi\fi\fi\@esphack}
        \gdef\@eqnlabel{#1}}
\def\@eqnlabel{}
\def\@vacuum{}
\def\draftmarginnote#1{\marginpar{\raggedright\scriptsize\tt#1}}

\def\draft{\oddsidemargin -.5truein
        \def\@oddfoot{\sl preliminary draft \hfil
        \rm\thepage\hfil\sl\today\quad\militarytime}
        \let\@evenfoot\@oddfoot \overfullrule 3pt
        \let\label=\draftlabel
        \let\marginnote=\draftmarginnote
   \def\@eqnnum{(\theequation)\rlap{\kern\marginparsep\tt\@eqnlabel}%
\global\let\@eqnlabel\@vacuum}  }

\def\draft2{
        \def\@oddfoot{\sl preliminary draft \hfil
        \rm\thepage\hfil\sl\today\quad\militarytime}
        \let\@evenfoot\@oddfoot \overfullrule 3pt
        \let\label=\draftlabel
        \let\marginnote=\draftmarginnote
   \def\@eqnnum{(\theequation)\rlap{\kern\marginparsep\tt\@eqnlabel}%
\global\let\@eqnlabel\@vacuum}  }

\def\preprint{\twocolumn\sloppy\flushbottom\parindent 2em
        \leftmargini 2em\leftmarginv .5em\leftmarginvi .5em
        \oddsidemargin -.5in    \evensidemargin -.5in
        \columnsep .4in \footheight 0pt
        \textwidth 10.in        \topmargin  -.4in
        \headheight 12pt \topskip .4in
        \textheight 6.9in \footskip 0pt
        \def\@oddhead{\thepage\hfil\addtocounter{page}{1}\thepage}
        \let\@evenhead\@oddhead \def\@oddfoot{} \def\@evenfoot{} }

\def\numberbysection{\@addtoreset{equation}{section}
        \def\theequation{\thesection.\arabic{equation}}}

\def\underline#1{\relax\ifmmode\@@underline#1\else
        $\@@underline{\hbox{#1}}$\relax\fi}

\def\titlepage{\@restonecolfalse\if@twocolumn\@restonecoltrue\onecolumn
     \else \newpage \fi \thispagestyle{empty}\c@page\z@
        \def\thefootnote{\fnsymbol{footnote}} }

\def\endtitlepage{\if@restonecol\twocolumn \else \newpage \fi
        \def\thefootnote{\arabic{footnote}}
        \setcounter{footnote}{0}}  

\catcode`@=12 \relax

\def\figcap{\section*{Figure Captions\markboth
        {FIGURECAPTIONS}{FIGURECAPTIONS}}\list
        {Figure \arabic{enumi}:\hfill}{\settowidth\labelwidth{Figure
999:}
        \leftmargin\labelwidth
        \advance\leftmargin\labelsep\usecounter{enumi}}}
 \relax
\def\tablecap{\section*{Table Captions\markboth
        {TABLECAPTIONS}{TABLECAPTIONS}}\list
        {Table \arabic{enumi}:\hfill}{\settowidth\labelwidth{Table
999:}
        \leftmargin\labelwidth
        \advance\leftmargin\labelsep\usecounter{enumi}}}
 \relax
\def\reflist{\section*{References\markboth
        {REFLIST}{REFLIST}}\list
        {[\arabic{enumi}]\hfill}{\settowidth\labelwidth{[999]}
        \leftmargin\labelwidth
        \advance\leftmargin\labelsep\usecounter{enumi}}}
 \relax
\makeatletter
\newcounter{pubctr}
\def\publist{\@ifnextchar[{\@publist}{\@@publist}}
\def\@publist[#1]{\list
        {[\arabic{pubctr}]\hfill}{\settowidth\labelwidth{[999]}
        \leftmargin\labelwidth
        \advance\leftmargin\labelsep
        \@nmbrlisttrue\def\@listctr{pubctr}
        \setcounter{pubctr}{#1}\addtocounter{pubctr}{-1}}}
\def\@@publist{\list
        {[\arabic{pubctr}]\hfill}{\settowidth\labelwidth{[999]}
        \leftmargin\labelwidth
        \advance\leftmargin\labelsep
        \@nmbrlisttrue\def\@listctr{pubctr}}}
 \relax
\makeatother

\def\be{\begin{equation}}
\def\ee{\end{equation}}
\def\ba{\begin{eqnarray}}
\def\ea{\end{eqnarray}}

\def\m{\mu}
\def\n{\nu}
\def\om{\omega}

\def\l{\lambda}

\def\s{\sigma}

\def\no{\noindent}

\def\IR{\relax{\rm I\kern-.18em R}}

\def\bse{\begin{small}\begin{equation*}}
\def\ese{\end{equation*}\end{small}}

\begin{document}


\renewcommand{\theequation}{\thesection.\arabic{equation}}
\csname @addtoreset\endcsname{equation}{section}

\newcommand{\eqn}[1]{(\ref{#1})}

\begin{titlepage}
\begin{center}
\strut\hfill
\vskip 1.3cm


\vskip .5in

{\Large \bf Scattering in Twisted Yangians}

\vskip 0.5in

 {\bf {Jean Avan$^{a}$, Anastasia Doikou$^{b}$\footnote{Based on a talk presented by AD, in {\it ``Integrable systems and quantum symmetries''}, Prague, June 2015. This work is mainly based on \cite{Avan:2014noa, Avan:2015}} and Nikos Karaiskos$^{c}$ } }\\
 \noindent
 \vskip 0.04in

{\footnotesize $^{a}$ Laboratoire de Physique Th\'eorique et Mod\'elisation
(CNRS UMR 8089), \\
Universit\'e de Cergy-Pontoise, F-95302 Cergy-Pontoise, France}\\[2mm]
\noindent
 \vskip 0.02in
{\footnotesize $^{b}$
 Department of Mathematics, Heriot-Watt University,\\
EH14 4AS, Edinburgh, United Kingdom}
\\[2mm]
\noindent

\vskip 0.02in
{\footnotesize $^{c}$ Institute for Theoretical Physics, Leibniz University
Hannover,\\ Appelstra\ss e 2, 30167 Hannover, Germany}
\\[2mm]
\noindent
\vskip .1cm


{\footnotesize {\tt E-mail: jean.avan@u-cery.fr, a.doikou@hw.ac.uk,
nikolaos.karaiskos@itp.uni-hannover.de}}\\

\end{center}

\vskip 1.0in

\centerline{\bf Abstract}

We study the bulk and boundary scattering of the $\mathfrak{sl}({\cal N})$ twisted Yangian spin chain via the solution of the Bethe ansatz equations in the thermodynamic limit. Explicit expressions for the scattering amplitudes are obtained and the factorization of the bulk scattering is shown. The issue of defects in twisted Yangians is also briefly discussed.

\no

\vfill

\end{titlepage}
\vfill \eject

\def\baselinestretch{1.2}

\tableofcontents

\section{Introduction}

We shall discuss here the bulk and boundary scattering in the context of twisted Yangians. Most of this material is described in more detail in \cite{Avan:2014noa, Avan:2015}. Here we are giving a brief review of the main results regarding basically the computation of the bulk and boundary scattering amplitudes in the case of the spin chain with twisted Yangian underlying algebra.

We shall deal henceforth with open spin chains, requiring introduction of boundary terms consistent with quantum integrability. These are related to generalized reflection algebras (quadratic algebras) \`a la Freidel-Maillet \cite{fm} extending the original construction of Cherednik \cite{cherednik} and Sklyanin \cite{sklyanin} to a four matrix structure canonically expressed as:
\be
A_{12}\ K_1\ B_{12}\ K_2 = K_2\ C_{12}\ K_1\ D_{12} \, ,
\label{quadr}
\ee
with unitarity requirements
\ba
&& A_{12}\ A_{21} = D_{12}\ D_{21} = {\mathbb I}_{12} \, ,\cr
&& C_{12}= B_{21}. \label{one}
\ea
In the particular case when $A_{12} =D_{21} =R_{12}$ a given Yang-Baxter $R$ matrix, and $B_{12} =C_{21} = \bar R_{21}$ (its soliton anti-soliton counter part), $\bar R_{12}\sim R_{12}^{t_1}$, (\ref{quadr}) yields the so-called twisted Yangian structure if $R$ is the simple Yangian solution of the Yang-Baxter equation \cite{Ols}.

Spin chains based on such a twisted Yangian were first constructed and investigated in \cite{Doikou:2000yw} for the first time whereas investigations were generalized in \cite{Arnaudon:2004sd}. They were then considered in the thermodynamic limit in our previous paper \cite{Avan:2014noa}. They naturally exhibit soliton non-preserving boundary conditions due the choice of $B_{12} = C_{21}$
as a soliton$-$anti-soliton $S$-matrix and the subsequent conversion of a soliton into an anti-soliton by the building reflection matrix $K$.

\section{Bethe ansatz equations in twisted Yangian}

Analytical Bethe ansatz techniques were applied  in \cite{Doikou:2000yw, Arnaudon:2004sd} to obtain the spectrum and BAEs for the twisted Yangian. Throughout the text we consider the boundary matrices, $c$-number representations of the twisted Yangian (\ref{one}) ($A_{12} =D_{21} =R_{12}$, and $B_{12} =C_{21} = \bar R_{21}$), to be proportional to unit. The spectrum of the $\mathfrak{sl}({\cal N})$ twisted Yangian is then given by the following expression:
\be
\Lambda(\lambda) = (a(\lambda)\ \bar b(\lambda))^{L}\ g_0(\lambda)\ A_0(\lambda) +
(b(\lambda)\ \bar b(\lambda))^{L} \prod_{j=1}^{{\cal N}-2}\ g_j(\lambda)\ A_j(\lambda) + (\bar a(\lambda)\ b(\lambda))^{L} g_{{\cal N}-1}(\lambda)\ A_{{\cal N}-1}(\lambda)
\ee
where we define:
\be
a(\lambda) = \lambda + i,
~~~~b(\lambda) = \lambda,
~~~~~\bar a(\lambda) = \lambda + i\rho -i,
~~~~\bar b(\lambda) = \lambda + i\rho
\ee
$g_l$ are terms due to boundary contributions
\ba
g_l(\lambda) &=& {\lambda + {i\rho \over 2} - {i\over 2} \over \lambda + {i\rho \over 2}} \cr
g_{{{\cal N} -1 \over 2}}(\lambda) &=& 1, ~~~~~{\cal N} ~~~~\mbox{odd} \cr
g_{{\cal N} -l +1}(\lambda) &=& g_{l}(-\lambda -i\rho) \cr
\rho &=& {{\cal N} \over 2}, ~~~~~~~ l \in \{1, \ldots {{\cal N}\over 2}-1 \}
\ea
and $A_l$ are the so called dressing functions defined as
\ba
A_0(\lambda) &=& \prod_{j=1}^{M^{(1)}} {\lambda + \lambda_j^{(1)} - {i \over 2} \over \lambda + \lambda_j^{(1)} + {i\over 2}} {\lambda + \lambda_j^{(1)} - {i \over 2} \over \lambda + \lambda_j^{(1)} + {i\over 2}}\cr
A_k(\lambda) &=&\prod_{j=1}^{M^{(k)}} {\lambda + \lambda_j^{(k)} - {i k \over 2} +i \over \lambda + \lambda_j^{(k)} + {ik\over 2}} {\lambda + \lambda_j^{(k)}  +{ik \over 2} +i \over \lambda + \lambda_j^{(k)} + {ik\over 2}} \cr
&\times & \prod_{j=1}^{M^{(k+1)}} {\lambda + \lambda_j^{(k+1)} +{ik\over 2}- {i \over 2} \over \lambda + \lambda_j^{(k+1)} + {i\over 2}} {\lambda - \lambda_j^{(k+1)} +{ik\over 2}- {i \over 2} \over \lambda - \lambda_j^{(k+1)} + {i\over 2}}\
\ea
\be
A_k(\lambda) = A_{{\cal N} - k +1}(-\lambda -i \rho), ~~~~k \in \{1, \ldots, {{\cal N} \over 2} - 1 \}
\ee
for ${\cal N} =2 n+1$
\ba
A_n(\lambda) &=& \prod_{j=1}^{M^{(n)}} {\lambda + \lambda_j^{(n)} + {i n \over 2} +i \over \lambda + \lambda_j^{(n)} + {in\over 2}}  {\lambda - \lambda_j^{(n)} + {i n \over 2} +i \over \lambda - \lambda_j^{(n)} + {in\over 2}} \cr &\times& {\lambda + \lambda_j^{(n)} + {i n \over 2} -{i\over 2} \over \lambda + \lambda_j^{(n)} + {i n \over 2}+ {i\over 2}} {\lambda - \lambda_j^{(n)} + {i n \over 2} -{i\over 2} \over \lambda - \lambda_j^{(n)} + {in\over 2}+ {i\over 2}}
\ea

Analyticity conditions imposed on the spectrum give rise to the associated Bethe ansatz equations presented below: defining
\be
e_n(\l) = \frac{\l+ \frac{in}{2}}{\l - \frac{in}{2}} \, ,
\ee
the BAE read as follows:
\begin{itemize}
\item  $\mathfrak{sl}(2n+1)$
\be
\begin{split}
e^L_{1}(\l_i^{(1)}) & = -
\prod_{j=1}^{M^{(1)}} e_{2}(\l_i^{(1)} - \l_j^{(1)})\,
e_{2}(\l_{i}^{(1)} + \l_j^{(1)})
\prod_{j=1}^{M^{(2)}}
e_{-1}(\l_i^{(1)} - \l_j^{(2)}) \,
e_{-1}(\l_i^{(1)} + \l_j^{(2)})\,, \cr
1 & = - \prod_{j=1}^{M^{(\ell)}}
e_{2}(\l_i^{(\ell)} - \l_j^{(\ell)}) \,
e_{2}(\l_i^{(\ell)} + \l_j^{(\ell)})
\prod_{ \tau = \pm1}
\prod_{ j=1}^{M^{(\ell+\tau)}} e_{-1}(\l_i^{(\ell)} - \l_j^{(\ell+\tau)}) \,
e_{-1}(\l_{i}^{(\ell)} + \l_{j}^{(\ell+\tau)}) \cr
& \hspace{0cm} \textrm{for} \qquad \ell = 2, \ldots, n-1, \cr
e_{-\frac{1}{2}}(\l_i^{(n)}) & = - \prod_{j=1}^{M^{(n)}}
e_{-1}(\l_i^{(n)} - \l_j^{(n)})\,
e_{-1}(\l_i^{(n)} + \l_j^{(n)}) \,
e_2(\l_i^{(n)} - \l_j^{(n)}) \,
e_2(\l_i^{(n)} + \l_j^{(n)}) \cr
& \qquad  \times  \prod_{ j=1}^{M^{(n-1)}}
e_{-1}(\l_i^{(n)} - \l_j^{(n-1)}) \,
e_{-1}(\l_i^{(n)} + \l_j^{(n-1)})  \, .
\end{split}
\label{BAE1b}
\ee
Note that in this case the Bethe ansatz equations are similar to the ones of the open $\mathfrak{osp}(1|2n)$ spin chain (see also \cite{Doikou:2000yw}, \cite{Arnaudon:2004sd}, \cite{Arnaudon:2003zw}).
\item $\mathfrak{sl}(2n)$
\begin{eqnarray}
  e_{1}^L(\l_{i}^{(1)}) &\!\!=\!\!& -\prod_{j=1}^{M^{(1)}}
  e_{2}(\l_{i}^{(1)} - \l_{j}^{(1)})\ e_{2}(\l_{i}^{(1)} +
  \l_{j}^{(1)})\ \prod_{ j=1}^{M^{(2)}}e_{-1}(\l_{i}^{(1)} -
  \l_{j}^{(2)})\ e_{-1}(\l_{i}^{(1)} + \l_{j}^{(2)})\,,
  \nonumber \\
  1 &\!\!=\!\!& - \prod_{j=1}^{M^{(\ell)}} e_{2}(\l_{i}^{(\ell)} -
  \l_{j}^{(\ell)})\ e_{2}(\l_{i}^{(\ell)} + \l_{j}^{(\ell)})\
  \prod_{\tau = \pm 1}\prod_{ j=1}^{M^{(\ell+\tau)}}
  e_{-1}(\l_{i}^{(\ell)} -
  \l_{j}^{(\ell+\tau)})\ e_{-1}(\l_{i}^{(\ell)} +
  \l_{j}^{(\ell+\tau)}) \nonumber \\
  && \ell= 2,\ldots,n-1, \nonumber \\
  e_{-1}(\l_{i}^{(n)}) &\!\!=\!\!& - \prod_{j=1}^{M^{(n)}}
  e_{2}(\l_{i}^{(n)} - \l_{j}^{(n)})\ e_{2}(\l_{i}^{(n)} +
  \l_{j}^{(n)}) \nonumber \\
  && \times \prod_{ j=1}^{M^{(n-1)}}e_{-1}^{2}(\l_{i}^{(n)} -
  \l_{j}^{(n-1)})\ e_{-1}^{2}(\l_{i}^{(n)} + \l_{j}^{(n-1)}).
\label{BAE2}
\end{eqnarray}
As opposed to the $\mathfrak{sl}(2n+1)$ case the Bethe ansatz equations above do not reduce to any of the known forms of BAE, which makes the whole study even more intriguing.

Note that the numbers $M^{(l)}$ are associated to the eigenvalues of the diagonal generators $S_l$ of the underlying algebra $\mathfrak{so}(n)$ (see \cite{Doikou:2000yw, Arnaudon:2004sd} for a detailed discussion on the underlying symmetry of the models), i.e.
\be
S_1 = {1\over 2} M^{(0)} - M^{(1)}, ~~~~S_l = M^{(l-1)} - M^{(l)}, ~~~~S_l = {1\over 2}(E_{ll} -E_{\bar l \bar l}), ~~~~ l\leq l \leq {{\cal N} -1 \over 2} \label{numbers}
\ee
$E_{ll}$ are the diagonal generators of $\mathfrak{sl}_{\cal N}$, and $\bar l = {\cal N} - l +1$ the conjugate index.

It is also worth recalling that the corresponding numbers in the usual $\mathfrak{sl}({\cal N})$ case are given by:
\be
E_{ll} = M^{(l-1)} -M^{(l)}, ~~~~M^{(0)} = 2L, ~~~~~M^{({\cal N})}=0,  ~~~~l \in \{1, 2, \ldots, {\cal N}\}
\ee
By imposing $M^{(l)} = M^{({\cal N} -1)}$ and considering the differences $E_{ll} - E_{\bar l \bar l}$  we end up with (\ref{numbers}) in accordance to the folding of $\mathfrak{sl}({\cal N})$ leading to the $\mathfrak{so}(n)$ algebra \cite{Doikou:2000yw, Arnaudon:2004sd}.
\end{itemize}

\section{Thermodynamics}

The aim now is to consider the study of the BAEs at the thermodynamic limit.
The ground state of the model consists of $n$ filled Dirac seas, unlike
the Yangian case, where the ground state consists of $2n+1$ or $2n$ filled
seas respectively.
As usual, an excitation corresponds to a hole in the Dirac sea. We perform
our computations in the thermodynamic limit of the BAE, which is obtained
according to the thermodynamic rule (for more details the interested reader is referred to e.g. \cite{Korepin:1979hg, Andrei:1983cb, Grisaru:1994ru} or \cite{Avan:2014noa} in a more relevant context)
\be
\frac{1}{L} \sum_{j=1}^{M^{(\ell)}} f(\l_j^{(\ell)}) \to
\int_0^{\infty} d\m \, \s_{\ell}(\m) \, f(\m) -
\frac{1}{L} \sum_{j=1}^{\n^{(\ell)}} f(\tilde{\l}_j^{(\ell)})
- \frac{1}{2L} f(0)\, ,
\ee
with $\n^{(\ell)}$ holes of rapidities $\tilde{\l}_j^{(\ell)}$ in the
$\ell^{\textrm{th}}$ Dirac sea $\s_{\ell}$ is the density in the $\ell^{th}$ sea. The last term is the halved contribution at $0^+$ due to the boundaries.
We shall focus here on the two-holes state, so that we can investigate both bulk and boundary scattering.
In the thermodynamic limit the densities describing the state in the presence of holes (particle-like excitations) are given as:
\be
\hat \sigma(\omega) = \hat \varepsilon(\omega) + {1 \over L}\ \hat r^{(1)}(\omega) \label{density}
\ee
where $\hat \sigma,\ \hat \varepsilon^{(0)},\ \hat r^{(i)}$ are $n$ column-vectors. In fact the $r^{(1)}$ contribution is the one that will provide the bulk and boundary scattering amplitudes as will be transparent in the subsequent section.

It is also worth noting that from the BAEs in the thermodynamic limit we can compute the energy of the holes in each sea:
\begin{itemize}
\item $\mathfrak{sl}(2n+1)$
\ba
\hat \varepsilon^{(j)}(\omega) = { \cosh(n+{1\over 2} -j){\omega \over 2} \over \cosh(n+{1\over 2} ){\omega \over 2}}, ~~~~j \in \{1, 2 \ldots, n\} \nonumber
\ea
\item $\mathfrak{sl}(2n)$
\ba
\hat \varepsilon^{(j)}(\omega) &=& { \cosh(n -j){\omega \over 2} \over \cosh  {n\omega \over 2}}, ~~~~j \in \{1, 2 \ldots, n-1\} \cr
\hat \varepsilon^{(n)}(\omega) &=& {1\over 2  \cosh  {n\omega \over 2}}
\ea
\end{itemize}
The details here are omitted but we refer the interested reader for more details in \cite{Avan:2014noa, Avan:2015}.

\section{Scattering}
The key element in this context is now the generalized quantization
condition for the twisted Yangian introduced in \cite{Avan:2014noa}. We shall
consider here the scattering of particle-like excitations in the first sea.
This is in fact inspired by earliest studies on the formulation of the quantization for quantum integrable systems with different boundary conditions (see also \cite{Korepin:1979hg},
\cite{Andrei:1983cb}, \cite{Grisaru:1994ru})

In the twisted Yangian a modified isomonodromy condition is imposed on the two-holes state \cite{Avan:2014noa, Avan:2015}:

\be
\Big(e^{i\mathcal{P}^{(\ell)}L} \,
\mathbb{S}(\tilde \l_1, \tilde \l_2) - 1 \Big)
| \tilde \l_1, \tilde \l_2\rangle = 0 \, ,
\label{mom_quant}
\ee
${\cal P}^{(\ell)}$ the momentum of the hole in the $\ell^{th}$ sea. The global scattering matrix ${\mathbb S}$ is given by:

\be
{\mathbb S}(\lambda_1,\ \lambda_2) =  {\cal K}^+(\lambda_1)\ {\cal S}(\lambda_1-\lambda_2)\
{\cal K}^-(\lambda_2)\ {\cal S}(\lambda_1+\lambda_2) \, ,
\ee
The ``bulk'' scattering ${\cal S}$ is factorized as
\be
{\cal S}(\lambda) = S(\lambda)\ \bar S(\lambda)
\ee
where $S$ is the soliton-soliton scattering matrix, and $\bar S$ is the soliton--antisoliton scattering matrix in the $\mathfrak{sl}({\cal N})$ spin chain.

The quantization condition is schematically depicted below:

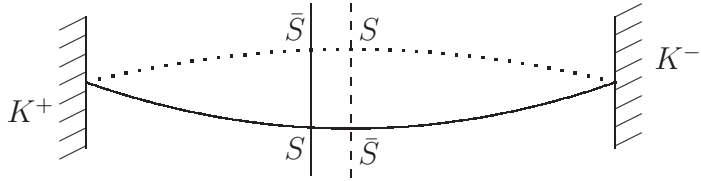
\begin{picture}(0,0)(-125,60)

\put(0,50){\line(0,-1){50}}
\multiput(0,50)(0,-7){8}{\line(-2,-1){10}}

\qbezier(0,25)(100,-10)(200,25)

\linethickness{0.3mm}

\qbezier[40](0,25)(100,50)(200,25)

\linethickness{0.1mm}

\put(200,50){\line(0,-1){50}}
\multiput(200,50)(0,-7){8}{\line(2,1){10}}

\put(85,55){\line(0,-1){65}}
\multiput(100,55)(0,-7){10}{\line(0,-1){3}}

\put(-30,10){$K^+$}
\put(75,41){$\bar{S}$}
\put(103,41){$S$}
\put(75,-4){$S$}
\put(103,-6){$\bar{S}$}
\put(215,30){$K^-$}

\end{picture}
\vskip 2.7cm

As already pointed out in the previous section we focus on the state with 2-holes in the first Dirac sea. If we now compare quantization condition with the density of the state (\ref{density}), recall also:
\be
\varepsilon^{(\ell)}(\lambda) = {i\over  2 \pi} {d {\cal P}^{(\ell)}(\lambda)\over d\lambda} \nonumber
\ee
then we conclude:
\be
\begin{split}
 \mathcal{S}_0(\l) & = \exp\Big \{-\int_{-\infty}^\infty \frac{d\om}{\om} \,
 e^{-i\om {\l}} \, \mathcal{B}_1(\om)\Big \} \cr
 K_0^+(\l) \, K_0^-(\l) & = \exp\Big \{- \int_{-\infty}^{\infty}
 \frac{d\om}{\om} \, \Big(
 e^{-i\om {\l}} \, \mathcal{B}_2(\om)
 + e^{-2i\om {\l}} \, \mathcal{B}_1(\om)
 \Big)\Big \} \, ,
\end{split} \label{scatter} \nonumber
\ee
where ${\cal S},\ K_0^{\pm}$ are eigenvalues of ${\cal S},\ {\cal K}^{\pm}$ respectively, and we define
\be
\begin{split}
\mathcal{B}_1(\om) & =
\hat{a}_2(\om) \, \hat{\mathcal{R}}_{11}(\om)
- \hat{a}_1(\om) \,  \hat{\mathcal{R}}_{12}(\om) \cr
\mathcal{B}_2(\om) & =
\sum_{j=1}^{n}
\big(
\hat{a}_2(\om) - 2\hat{a}_1(\om) + \hat{a}_1(\om) \delta_{i1} -
\hat{a}_{\frac{1}{2}}(\om) \delta_{in}
\big)
\hat{\mathcal{R}}_{1i}(\om) \, .
\end{split}
\label{bulk_bound}
\ee
\be
\hat a_n(\omega) = e^{-{n|\omega|\over 2}}, ~~~~\hat{\mathcal{R}}_{ij}(\omega) = e^{\omega \over 2} {\sinh \Big ( \mbox{min(i, j)}{\omega \over 2} \Big )\ \cosh\Big (n+{1\over 2}-\mbox{max}(i, j) \Big ){\omega \over 2} \over \cosh\Big (n+{1\over 2} \Big ){\omega \over 2}\ \sinh {\omega \over 2} }.
\ee
Explicit expressions for bulk and boundary scattering amplitudes are manifestly extracted above and the bulk scattering factorization: ${\cal S}(\lambda) = S(\lambda)\ \bar S(\lambda)$ may be then shown (we refer the interested reader to \cite{Avan:2014noa, Avan:2015} for more details).

\section{Implementing defects}
We shall briefly discuss here the case where a local defect is implemented.
\be
\mbox{Define}: ~~~~ X_k^+(\lambda)= {\lambda + i \alpha_k -{ik\over 2} \over \lambda + i \alpha_{k+1} -{ik\over 2} }, ~~~~~X_k^-(\lambda)= {\lambda + i \alpha_{{\cal N}-k+1} +{i ({\cal N}-k)\over 2} \over \lambda + i \alpha_{{\cal N}-k} +{i ({\cal N}-k)\over 2} } \nonumber
\ee
then the BAEs via the analytical Bethe ansatz formulation read as
\begin{itemize}
\item $\mathfrak{sl}(2n+1)$
\end{itemize}
\ba
&& X_1^+(\lambda_i^{(1)} -\Theta)\ X_1^+(\lambda_i^{(1)} +\Theta)\ e^L_{1}(\l_i^{(1)}) =  \cr
&& \qquad -
\prod_{j=1}^{M^{(1)}} e_{2}(\l_i^{(1)} - \l_j^{(1)})\,
e_{2}(\l_{i}^{(1)} + \l_j^{(1)})
\prod_{j=1}^{M^{(2)}}
e_{-1}(\l_i^{(1)} - \l_j^{(2)}) \,
e_{-1}(\l_i^{(1)} + \l_j^{(2)})\,, \cr
&& X_{\ell}^+(\lambda_i^{(l)} -\Theta)\ X_{\ell}^+(\lambda_i^{(l)} +\Theta) =\cr
&& \qquad - \prod_{j=1}^{M^{(\ell)}}
e_{2}(\l_i^{(\ell)} - \l_j^{(\ell)}) \,
e_{2}(\l_i^{(\ell)} + \l_j^{(\ell)})
\prod_{ \tau = \pm1}
\prod_{ j=1}^{M^{(\ell+\tau)}} e_{-1}(\l_i^{(\ell)} - \l_j^{(\ell+\tau)}) \,
e_{-1}(\l_{i}^{(\ell)} + \l_{j}^{(\ell+\tau)}) \cr
&& \qquad \textrm{for} \qquad \ell = 2, \ldots, n-1, \cr
&& X_n^+(\lambda_i^{(n)} -\Theta)\ X_n^+(\lambda_i^{(n)} +\Theta)\ e_{-\frac{1}{2}}(\l_i^{(n)})
= - \prod_{ j=1}^{M^{(n-1)}}
e_{-1}(\l_i^{(n)} - \l_j^{(n-1)}) \,
e_{-1}(\l_i^{(n)} + \l_j^{(n-1)})  \,  \cr
&& \qquad \times - \prod_{j=1}^{M^{(n)}}
e_{-1}(\l_i^{(n)} - \l_j^{(n)})\,
e_{-1}(\l_i^{(n)} + \l_j^{(n)}) \,
e_2(\l_i^{(n)} - \l_j^{(n)}) \,
e_2(\l_i^{(n)} + \l_j^{(n)})
\ea

Notice that the Bethe ansatz equations in the presence of defects are very similar to the ones presented in the previous section. The main difference is the existence of the extra contributions $X^{\pm}$ due to the presence of defects. In the thermodynamic limit these contributions will provide the transmission amplitudes.

We  may now formulate a suitable
quantization condition for the model in the presence of defects. In order to
determine the relevant transmission matrix it suffices to consider a state with
one hole in the first sea. Let us first introduce some notation and define the
transmission amplitudes in $\mathfrak{sl}({\cal N})$ \cite{Doikou-trans} as
\ba
T(\lambda-\Theta):&& \mbox{soliton$-$defect scattering}\cr
\bar T(\lambda-\Theta):&& \mbox{soliton$-$anti-defect scattering}\cr
T^*(\lambda+\Theta):&& \mbox{anti-soliton$-$defect scattering}\cr
\bar T^*(\lambda+\Theta):&& \mbox{anti-soliton$-$anti-defect scattering}
\ea

The quantization condition for such a state reads as
\be
\Big (e^{i{\cal P}^{(l)}}{\mathbb S}(\tilde \l^{(l)}, \Theta) - 1\Big )
|\tilde \l^{(l)}, \Theta \rangle =0 \, ,
\ee
where the global scattering amplitude is given by
\be
{\mathbb S}(\lambda, \Theta) = {\cal K}^+(\lambda)\ T(\l - \Theta)\ \bar T(\l - \Theta)\
{\cal K}^-(\lambda)\  \bar T^*(\l +\Theta)\ T^*(\l +\Theta) \label{kt}
\ee
We shall not give further details on the derivation of transmission amplitudes in twisted Yangians. Explicit expressions of transmission amplitudes and their factorizations are provided in \cite{Avan:2015}. With this we conclude our presentation on the bulk and boundary scattering in the context of twisted Yangians.

\end{document}